\documentclass[
twocolumn,superscriptaddress,
amsmath,amssymb,aps,prfluids,floatfix,
]{revtex4-2}

\usepackage{hyperref} 
\usepackage{verbatim}
\usepackage{graphicx}
\usepackage{dcolumn}
\usepackage{bm} 
\usepackage{ifpdf} 
\usepackage{dcolumn}  
\usepackage{epsfig}
\usepackage{epstopdf} 
\usepackage{color}

\begin{document}
	
	\title{Activity-induced asymmetric dispersion in confined channels with constriction}
	\author{Armin Maleki}
	\author{Malihe Ghodrat}
	\email{m.ghodrat@modares.ac.ir}
	\affiliation{Department of Physics, Faculty of Basic Sciences, Tarbiat Modares University, P.O. Box 14115-175, Tehran, Iran}
	\author{Ignacio Pagonabarraga}
	\email{ipagonabarraga@ub.edu}	
	\affiliation{Departament de F\'{\i}sica de la Mat\`eria Condensada, Universitat de Barcelona, Carrer de Mart\'{\i} i Franqu\'es 1, 08028 Barcelona, Spain}
	\affiliation{  Universitat de Barcelona Institute of Complex Systems (UBICS), Universitat de Barcelona, 08028 Barcelona, Spain}
	
	%
	\begin{abstract}
		Microorganisms, such as E.Coli, are known to  display upstream behavior and respond rheotactically to  shear flows. In particular, E.Coli suspensions have been shown to   display strong sensitivity to   spatial constrictions, leading to  an anomalous densification past the constriction for incoming fluid velocities comparable to the microoganism's self propulsion speed. We introduce a Brownian dynamics model for ellipsoidal self-propelling particles  in a confined channel subject to a constriction. The model allows to identify the relevant parameters that characterize the relevant dynamical regimes of  the accumulation of  the  active particles at the constriction, and clarify the mechanisms underlying the experimental observations. We find that particles are trapped in butterfly-like attractors in front of the constriction, which  is the origin of the symmetry breaking in the emerging density profiles of active particles passing the constriction. In addition, the probability of trapping and thus the strength of asymmetry is affected by size of the particles and geometry of the channel, as well as the ratio of fluid velocity to propulsion speed.
	\end{abstract}
	
	\maketitle
	

	\section{Introduction}
	Microorganisms  can be found in a wide variety of  media and complex environments~\cite{Bechinger2016}. Their emergent, collective behavior, is a  result of the interrelated motion  due to the incoming flow features  and the  disturbance microorganisms generate. Such hydrodynamic coupling    significantly alters the rheological properties of active suspensions~\cite{Dey2022},  modifies how  microorganisms swim and disperse, as well as strongly  impacts the regimes of solute transport~\cite{Rusconi2014}.  This  leads to a wide and rich variety of emerging behavior, which includes anomalous viscosity~\cite{prl_clement}, mixing  enhancement, bioconvection or anomalous dispersion~\cite{Creppy2019}.  In the presence of a shear flow, they show rheotactic behavior~\cite{Jing2020}, which depends on the  microorganism shape and mode of locomotion~\cite{Daddi2020}, e.g. for flagellates, the interaction of the moving flagella with the shear flow is known to play  a critical role.  Positive rheotaxis, i.e. upstream navigation, has been observed under confinement both for a variety of  microorganisms, such as  sperm cells in the reproductive tract~\cite{Lane2005},  bacteria in the upper urinary tract, and  E. coli in catheters~\cite{Figueroa2020}.
	
	In particular, when microorganisms swim close to  confining walls, their behavior can be qualitatively altered, leading to   chiral trajectories and  upstream motion~\cite{zottl2012}. Microorganisms, in such conditions, show  a high sensitivity to  morphological variabilities of the confining substrate~\cite{Schmidt2022}. Such coupling and sensitivity has strong implications in  the behavior of microorganisms in porous media, how they are transported or dispersed,  or how they organize and accumulate inside such heterogeneous media~\cite{Saintillan2019, Strump2021}. 
	
	The collective behavior of microswimmers in confined channels and in the presence of constrictions has significant implications for the transport and dispersion of bacteria in fluid environments~\cite{Liu2011}. These tiny organisms have constitutive properties that differ significantly from passive suspensions, leading to new and surprising effects such as activated Brownian motion, anomalous viscosity, mixing enhancement, bioconvection, and work extraction from fluctuations.
	
	Despite the practical implications for bio-contamination in porous rocks, biological micro-vessels, and medical catheters, the fundamental question of hydrodynamic dispersion of bacteria suspended in a fluid remains a challenge~\cite{Dentz2022}. Current methods of analysis rely on macroscopic convection-diffusion equations with adsorption-desorption terms to describe retention effects by surfaces \cite{clement2013a}. However, systematic inconsistencies between experiment and modeling suggest the need for further refinements, including the detailed interaction of individual microorganisms and the confining substrate.  Studies on simplified geometries and model pores provide  systematic analysis and understanding of the response of microorganisms under confinement. For example, recent experiments~\cite{clement2013} have quantified the dispersion of  E. Coli through a funnel,  showing an anomalous downstream densification of the microorganisms for  incoming fluid velocities comparable to the self propulsion speed of E. Coli.
	
	In this work we will introduce a simple, general theoretical model that describes the motion of self-propelling particles (SPPs) in a channel characterized by a constriction, subject to an incoming   fluid flow.  The  model allows to identify the relevant dimensionless parameters that  characterize how SPPs accumulate  around the constriction and helps to identify the physical mechanisms that control the motion and organization of SPPs around a constriction. In Section \ref{sec:model} we introduce the model and details of the geometry and the methodology to  solve it. Section \ref{sec:dimensionless} identifies the relevant dimensionless parameters that identify the relevant dynamical regimes of SPPs suspensions. Subsequently,  Section \ref{sec:results} systematically presents the relevant quantities that characterize the  emerging behavior of SPPs  in a confined channel in the presence  of a constriction. The manuscript finishes  highlighting the main results obtained and their  implications in Section \ref{sec:conclusions}.
	
	\section{Model and simulation techniques}
	\label{sec:model}
	\begin{figure}
		\begin{center}
			\begin{minipage}[h]{\linewidth}	\begin{center}
					\includegraphics[width=\linewidth]{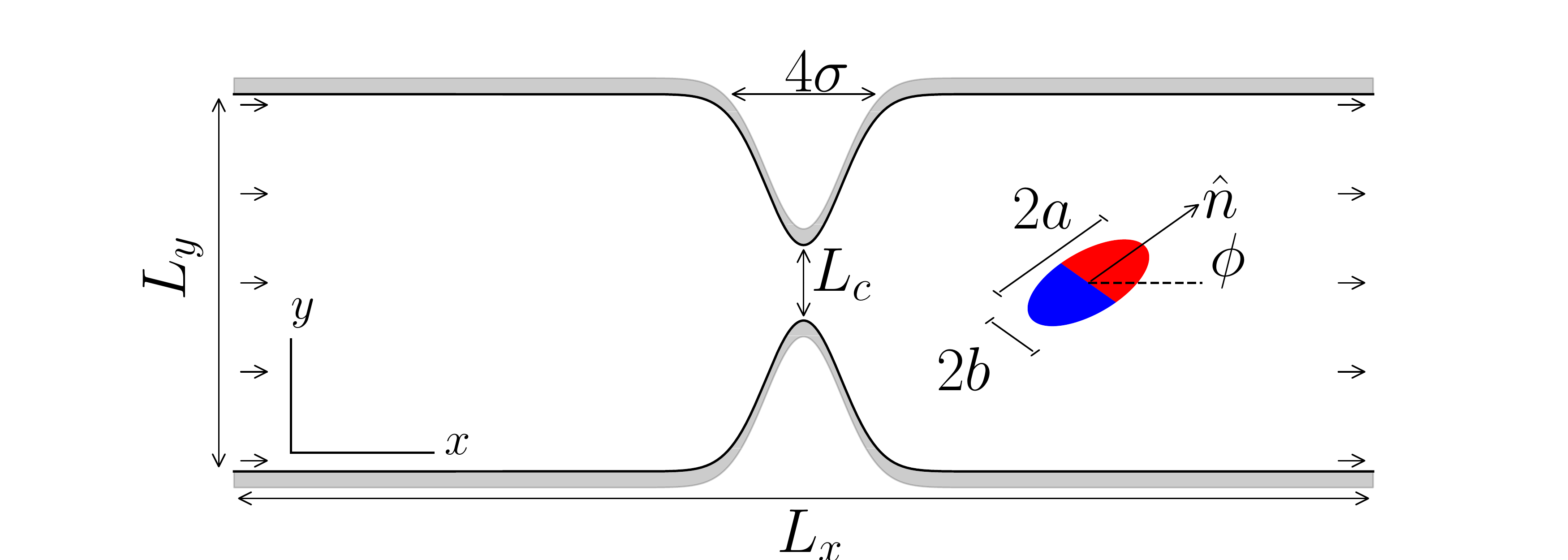} \\ (a) 
			\end{center} \end{minipage} \vskip0.2cm
			\begin{minipage}[h]{\linewidth}	\begin{center}
					\includegraphics[width=0.9\linewidth]{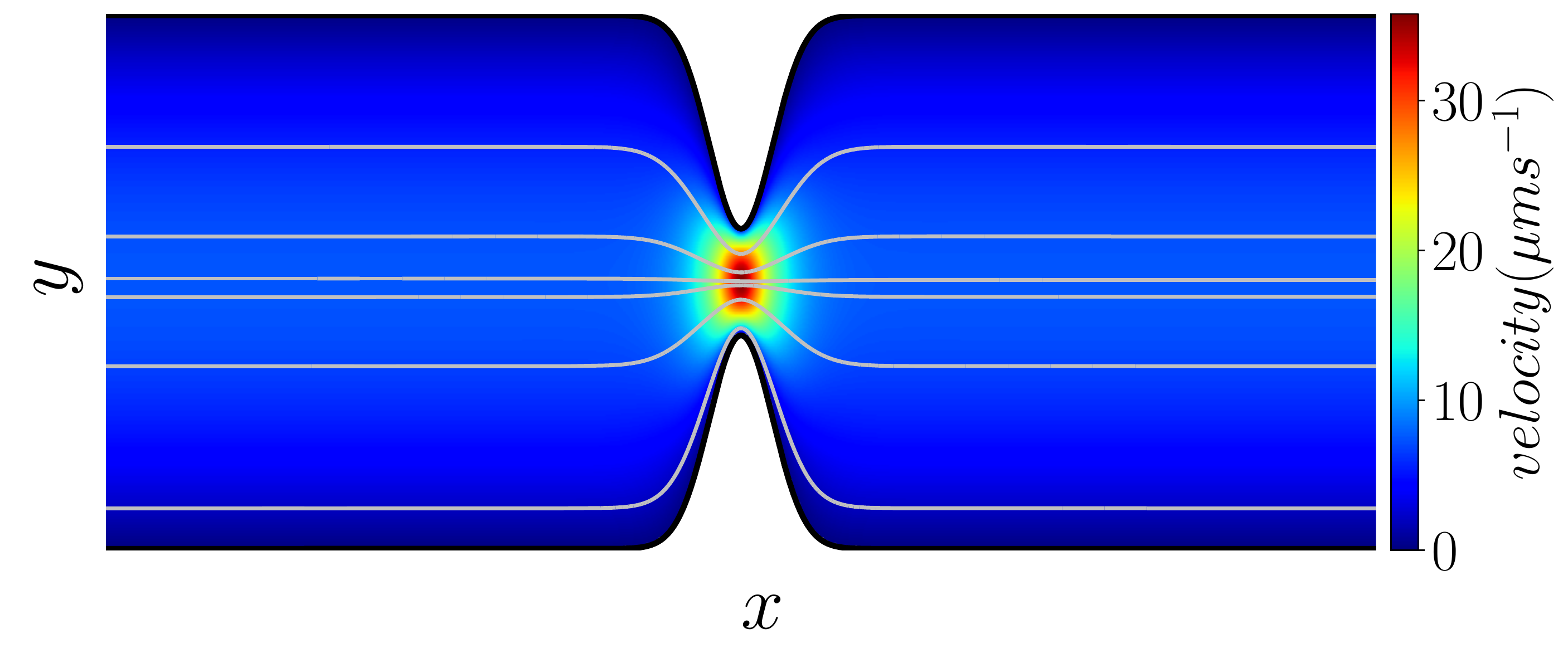} \\(b) 
			\end{center} \end{minipage} 
		\end{center}
		\caption{(a) Schematic of the  system used in this study, displaying the relevant 
			geometry and parameters. (b) Velocity field and streamlines in the channel of $L_x=4mm$, $L_y=200\mu m$, $L_c=40\mu m$ and incoming velocity $ u_{in} = 5.0\,\mu m s^{-1}$ corresponding to an average velocity $\langle u_x \rangle= 5.1\, \mu m s^{-1}$.}
		\label{f:schematic}
	\end{figure}
	The model system consists of $N$ SPPs swimming in a rectangular channel of length $L_x$ and height $L_y$, with a constriction  described by a Gaussian  
	\begin{equation}
		f(x)= h \exp(-\frac{(x - x_{mid}) ^2}{2 \sigma ^2}),
	\end{equation}
	with height $h=80 \mu m$, expanded overa  length $\sigma=50 \mu m$ (for the lower boundary), as shown in Fig.~\ref{f:schematic}.a. In the centering point ($x_{mid}=0$), the channel reaches the minimum width of $L_c=L_y-2h$ which varies between $15-240\,\mu m$ in our simulations. 
	The suspending media in the channel is a Newtonian fluid of density $\rho=1000 Kgm^{-3}$ and dynamic viscosity $\eta=0.88\,mPas$ at temperature $T=298K$. The size of SPPs is typically $1-10\,\mu m$ and their self-propulsion velocity, $v_s\sim \mathcal O (10)\,\mu m s^{-1}$. 
	
	This velocity induces low Reynolds numbers flows for spherical particles of radius $a$~\cite{rhodes2008}, $$Re_s=\frac{\rho v_s a}{\eta} \sim10^{-5},$$
	in which the inertial forces are negligible compared to the viscous forces that swimming objects experience in the fluid~\cite{purcell1977}. In the following we present the details of our assumptions for suspending fluid, swimming particles and boundaries.  
	
	\paragraph{Fluid--}The suspending fluid enters the left side of the channel with incoming velocity $u_{in}$ due to a pressure gradient, and forms a laminar flow with typical average velocity $\langle u \rangle \sim \mathcal O (10)\,\mu m s^{-1}$, such that the flow can also be considered as a laminar one  in low Reynolds regime \cite{sommerfield1908}, $$ Re_f=\frac{\rho \langle u\rangle L_x}{\eta} \sim10^{-2}$$ 
	
	The fluid velocity field in the channel, $\mathbf{u(r)}$, is obtained by solving the linear form of Navier-Stockes equation, which reduces to the Stokes equation at vanishingly Reynolds numbers
	\begin{equation} 
		\label{eq:NS1}
		\eta \nabla^2 \mathbf u=\nabla P
	\end{equation}
	The fluid is assumed to be an incompressible Newtonian fluid, $\nabla.\mathbf u=0$, and $\nabla P$ denotes the pressure gradient \cite{kirby2010}. Eq.~(\ref{eq:NS1}) is solved numerically by finite element method with no slip boundary  condition on the channel walls. Fig.\ref{f:schematic}.b indicates the velocity field and streamlines for a channel with incoming velocity $u_{in} =5.0 \mu ms^{-1}$, and funnel width, $L_c=40\mu m$.  The fluid velocity attains its maximum value at the center of the constriction and goes to zero in the vicinity of the walls, as expected. Additionally, due to continuity and incompressibility, the average velocity in the channel is linearly proportional to the incoming velocity (or equivalently the fluid velocity far from the constriction) for arbitrary channel widths; i.e.,  
	$$\frac{\langle u_x\rangle}{u_{in}}\propto \frac{L_y}{L_c}.$$

	\paragraph{Particles--}The motility of spherical and/or elliptical particles, swimming in the channel, are modeled by two dimensional over-damped Langevin equations
	\begin{eqnarray}
		\dot{\mathbf r} &=& v_s \hat{\mathbf n} + \mathbf{u (r)}+ \sqrt{2 D_T}\;\xi_r \label{e:Br_Dynm_1}\\
		\dot\phi&=& \Omega_0+\Omega_{f}(\mathbf r)+\sqrt{2 D_R}\;\xi_\phi,
		\label{e:Br_Dynm_2}
	\end{eqnarray}
	in which $\hat{\mathbf n}=(\cos\phi,\sin\phi)$ indicates the particle polarity  and $\phi$ 
	denotes the angle in polar coordinates. In this model, which is best known as active Brownian particle (ABP) model, $D_T$ and $D_R$ represent the translational and rotational diffusion coefficients, respectively. For spherical particles of radius $a$, the rotational diffusion is $D_R=k_B T/8 \pi \eta a^3$ and $D_T=4a^2 D_R/3$. The corresponding expressions for ellipsoidal particles are given in \cite{ellipse_diff}.
	
	The noise term $\xi$, is a zero averaged, $\langle \xi \rangle=0$, uncorrelated, $\langle \xi_\alpha(t)\xi_\beta(t') \rangle=\delta_{\alpha\beta}\delta(t-t')$, random process described by  a Gaussian distribution, which simulates the thermal fluctuations experienced by the particles in their translational ($\xi_r$) and rotational ($\xi_\phi$) motion. 
	
	The self-propelling term, $v_0 \hat{\mathbf n}$, provides a directed motion which couples the translational and rotational degrees of freedom through the polarity vector. The term, $\mathbf{u(r)}$, accounts for the background fluid velocity. In the dilute regime, which is the situation to be considered in this work, it can be assumed that the fluid velocity (as well as the fluid viscosity) is not affected in the presence of self-propelling particles. Therefore, the fluid velocity obtained from Eq.~\ref{eq:NS1} is taken as a time independent quantity in the Langevin equation. 
	
	On the other hand, according to Jeffery's equation\cite{jeff}, the spherical and/or ellipsoidal particles experience a torque due to the non-uniform field of the fluid velocity that applies asymmetric forces on the upper and lower parts of the propelling objects. The resulting angular velocity, $\Omega_f(\mathbf r)$,  for 2D ellipsoidal particles read
	\begin{equation}
		\Omega_{f}(\mathbf r)=\frac{1}{2}[u_{yx}-u_{xy}]+\frac{\beta}{2}[u_{yx}+u_{xy}]\cos(2\phi)-\beta \sin(2\phi),
		\label{e:omg_ellips}
	\end{equation}
	where $\beta=(\lambda^2-1)/(\lambda^2+1)$ defines the shape eccentricity, $\lambda=a/b$, being the ratio of semi-major to semi-minor axis of ellipsoids (or aspect ratio) and $u_{xy}=\partial u_x/\partial y$. In the special case of spherical particles ($\beta=0$), the above equation reduces to
	\begin{equation}
		\Omega_{f}(\mathbf r)=\frac{1}{2}[u_{yx}-u_{xy}].
		\label{e:omg_sph}
	\end{equation}
	
	The first term in Eq.~\ref{e:Br_Dynm_2}, considers the intrinsic angular velocity,  $\Omega_0$, caused by chirality, which is zero for the symmetric spherical/ellipsoidal particles used in this study. We have neglected the inter-particle interactions since we consider dilute suspensions. The system is not subject to external forces.

	\paragraph{Boundaries--}The upper and lower boundaries are solid walls which return the colliding particle back to the channel, based on a mirror reflection rule, from the tangential line at the colliding point~\cite{volpe2014}. On the side walls, we consider periodic boundary conditions; a particle that exits one side of the channel enters the opposite side while its vertical position and propelling direction are kept unchanged. Hence, we   conserve particle number  and minimize system size effects. 
	
	\section{Dimensionless equations }
	\label{sec:dimensionless}
	Using the inverse of  the rotational diffusion constant as time scale, $\tau_p=D_R^{-1}$, and the characteristic size of the SPPs, $a$, as reference length scale, we may rewrite Eqs.~\ref{e:Br_Dynm_1},\ref{e:Br_Dynm_2} in dimensionless form:
	\begin{eqnarray}
		\label{e:DL_Br_Dynm_21} \hat {\dot{ \mathbf r}} &=& Pe_s \hat{\mathbf n} + Pe_f \mathbf{\hat u (\hat r)}+ \sqrt{2 \hat D_T}\;\hat \xi_r \label{e:DL_Br_Dynm_1}\\
		\label{e:DL_Br_Dynm_22}\hat {\dot\phi}&=& \hat\Omega_0+\hat \Omega_{f}(\mathbf r)+\sqrt{2}\;\hat \xi_\phi
	\end{eqnarray}
	where we have chosen the incoming fluid velocity, $u_{in}$, as characteristic velocity of the fluid flow \cite{Note1}, such that $\mathbf{\hat u (\hat r)}=\mathbf{u (r)}/ u_{in}$, $\hat{\Omega}(\hat{{\bf r}})=D_R^{-1}\Omega({\bf r})$,  $\tilde \xi = D_R^{-1/2} \xi$, and the dimensionless parameter $$\hat D_T= \frac{D_T}{a^2 D_R}$$ is the ratio of the rotational diffusion time scale, $D_R^{-1}$ to that of translational diffusion, $a^2/D_T$. Other relevant dimensionless parameters are  SPPs' P\'{e}clet number, $$Pe_s=\frac{v_s}{a D_R},$$ and the fluid's P\'{e}clet number, $$Pe_f=\frac{u_{in}}{a   D_R}=\frac{u_{in}}{v_s} Pe_s.$$ The latter compares the ratio of diffusive time $D_R^{-1}$, to convection time $a/u_{in}$ or equivalently the ratio of convective length $u_{in}/D_R$ to the characteristic length scale, $a$. 
	
	Eqs.~\ref{e:DL_Br_Dynm_21}, \ref{e:DL_Br_Dynm_22} show that the system is determined by  eight dimensionless parameters: the active particle P\'eclet number, $Pe_s$, the fluid P\'eclet number $Pe_f$, the active vorticity $\hat{\Omega}_0$ and dimensionless fluid vorticity $\hat{\Omega}_{f}$, the  relative magnitude of the translational diffusion, $\hat {D}_T$, the SPP number density, $\rho$, the relative size of the channel constriction, $L_c/L_y$, and the relative channel width with respect to the size of microswimmer, $\hat{L}_y\equiv L_y/a$. The    persistent length, $l_p=v_s D_R^{-1}$ is also a relevant length scale in active systems. We will use the dimensionless parameter $\tilde X\equiv X/l_p$ whenever a length scale is compared with persistent length.
	
	We will consider achiral SPPs, hence $\hat{\Omega}_0$=0, and disregard the interactions among SPPs, hence $\rho$ is not relevant (except for being related to the number of particles simulated, which determine the statistics of the numerical simulations). We will also fix the ratio between the translational and rotational diffusion coefficient. Therefore, we are left with 5 relevant parameters: two determine the active regime of the SPPs under an incoming fluid flow, and three geometric ones, related to the degree of confinement of the SPPs and the particle asymmetry:  $Pe_s$, $Pe_f$, $L_y/a$, $L_c/L_y$ and $\lambda$.  
	
	\section{Results}
	\label{sec:results}
	Unless otherwise stated, SPPs are ellipsoids of semi-minor axis, $b=0.5 \mu m$, and aspect ratio, $\lambda=a/b=6$, which propel in the channel described in Fig.~\ref{f:schematic} with constant speed $v_s=20\mu m s^{-1}$,  and rotational diffusion $\tau_p=D_R^{-1}=6.85s$. which corresponds to a  persistent length  $l_p=v_s \tau_p=137\mu m$, and $Pe_s=45.7$. We will also fix the channel width, $a/L_y=3/200$ and the constriction size, $L_c/L_y=40/200$.  We will hence  analyze the  relevant regimes varying the relative strength of the incoming flow, $Pe_f/Pe_s=u_{in}/v_s\equiv\tilde u_{in}$, and the SPP asymmetry, $\lambda$, and will also consider the impact of the microswimmer persistent length varying $l_p/L_y$. 
	
	\subsection{Trajectories}
	\begin{figure}[t!]
		\begin{center}
			\begin{minipage}[h]{0.49\linewidth}	\begin{center}
					\includegraphics[width=\linewidth]{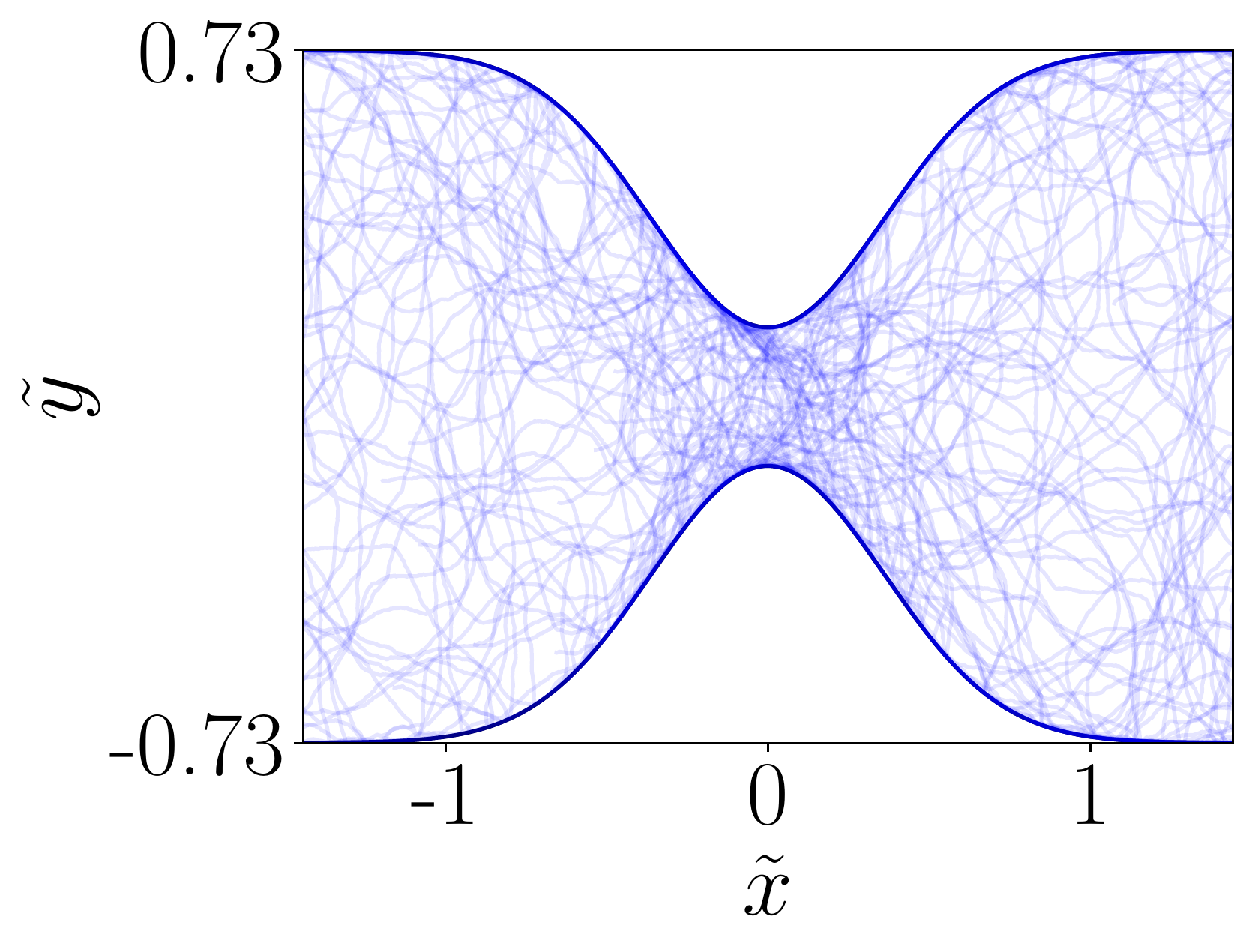}  (a) $\tilde u_{in}=0$
			\end{center} \end{minipage} \hskip0.01cm	
			\begin{minipage}[h]{0.49\linewidth}	\begin{center}
					\includegraphics[width=\linewidth]{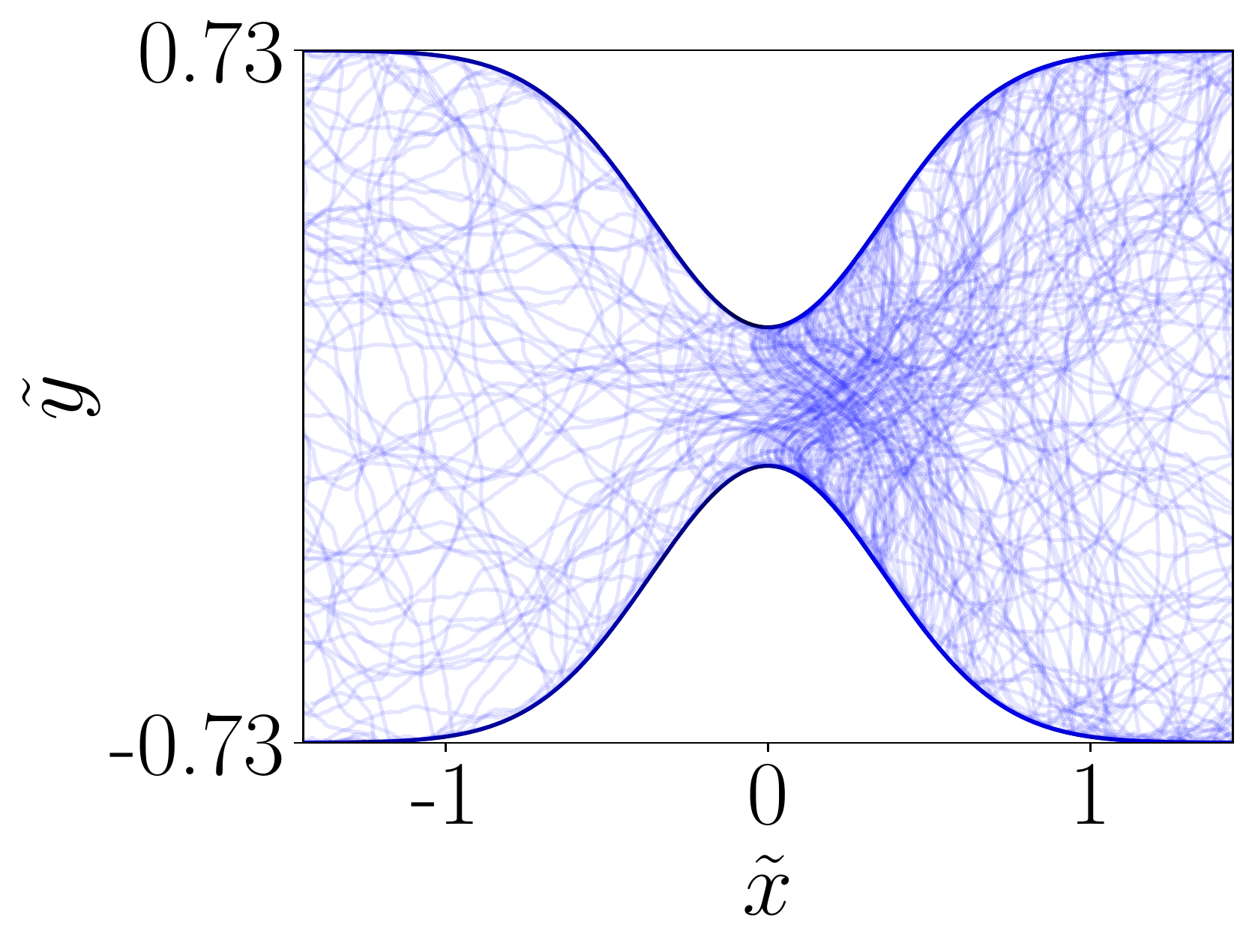} (b) $\tilde u_{in}=1/4$
			\end{center} \end{minipage} \hskip0.01cm\\
			\begin{minipage}[h]{0.49\linewidth}	\begin{center}
					\includegraphics[width=\linewidth]{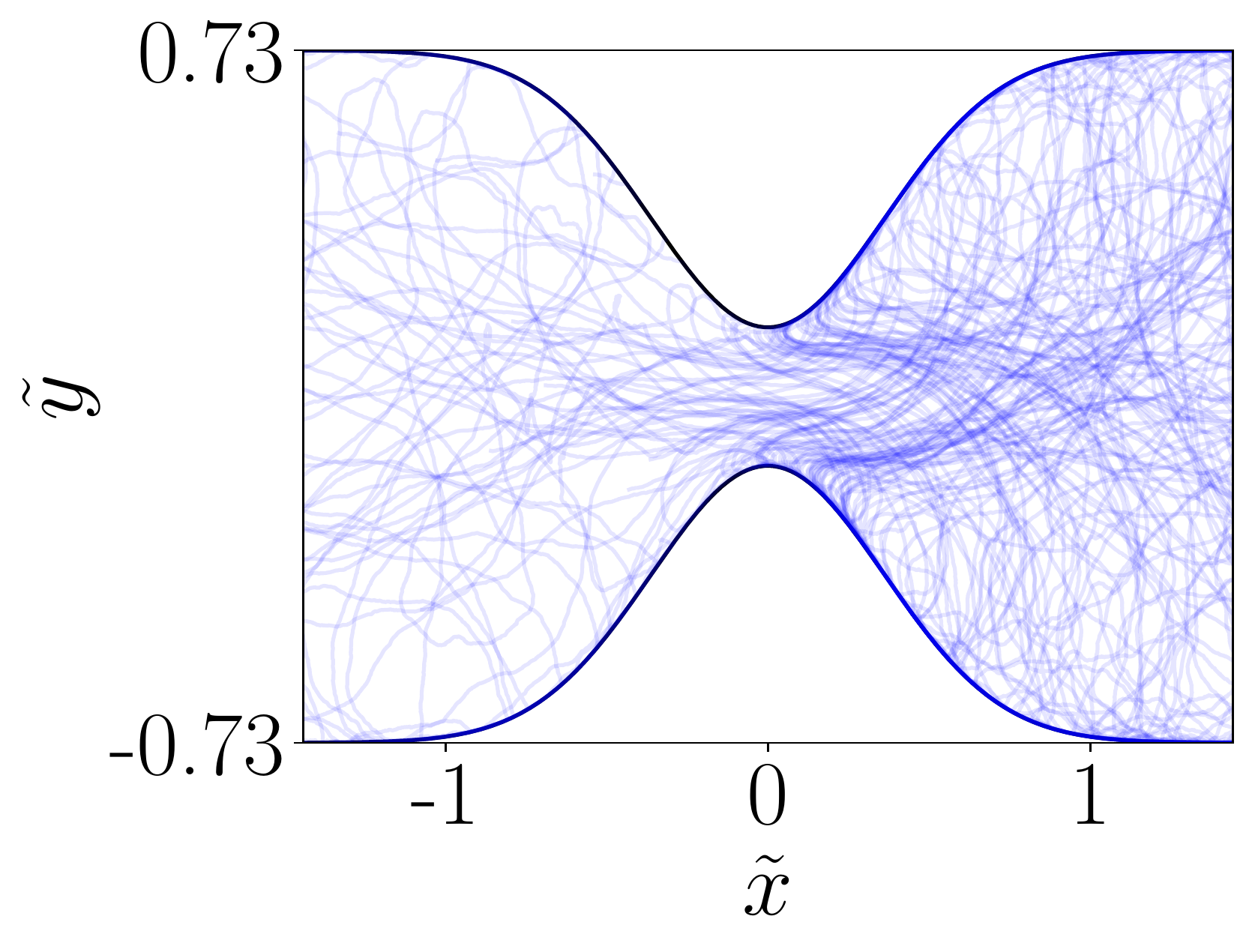} (c) $\tilde u_{in}=1/2$
			\end{center} \end{minipage} \hskip0.01cm	
			\begin{minipage}[h]{0.49\linewidth}	\begin{center}
					\includegraphics[width=\linewidth]{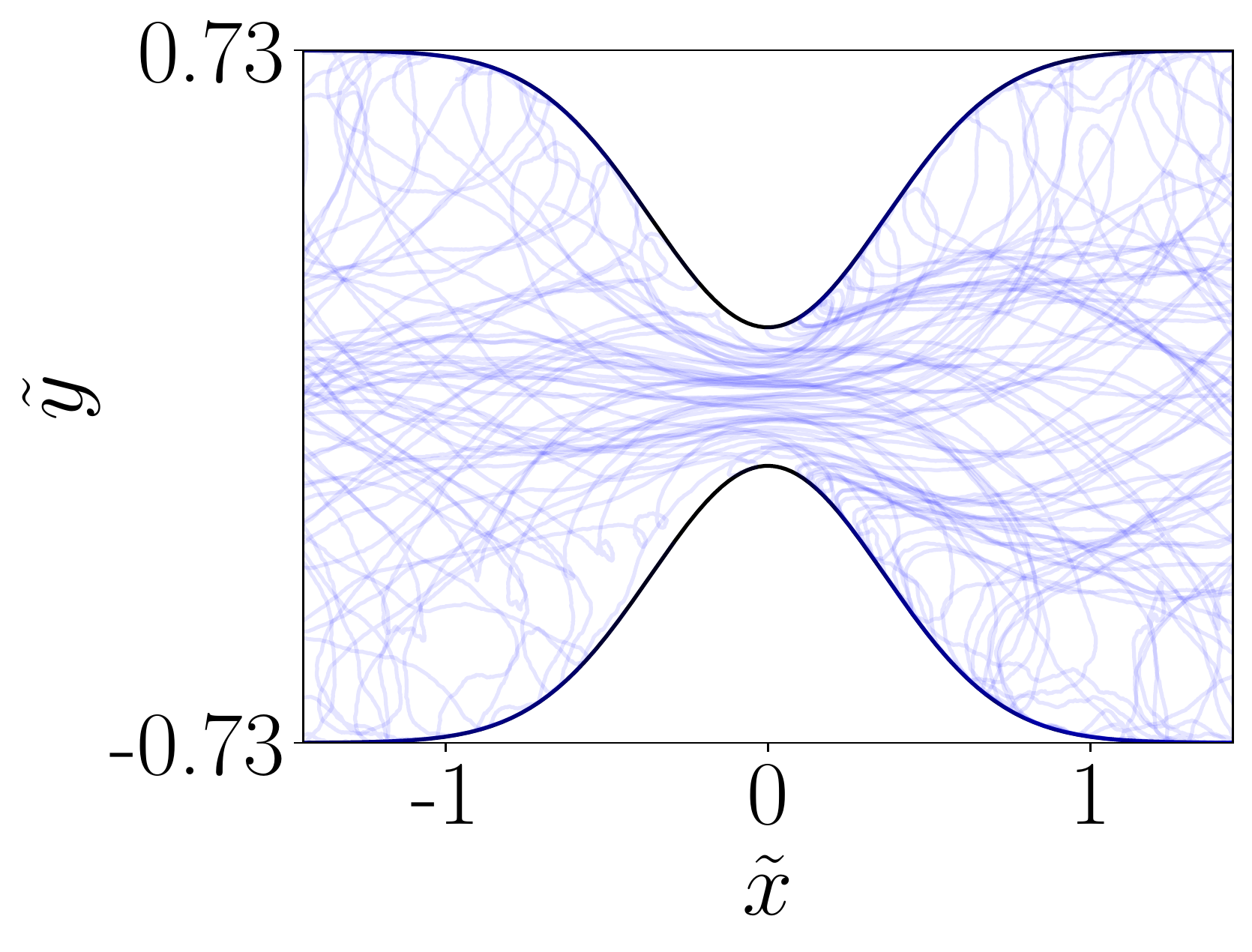} (d) $\tilde u_{in}=1$
			\end{center} \end{minipage} \hskip0.01cm	
		\end{center}
		\caption{Trajectories of 50 ellipsoidal  particles of aspect ratio $\lambda=6$ and $Pe_s=45.7$ ($v_s=20 \mu m s^{-1}$) within a time interval of $t\sim30\tau_p$ . Panels (a-d) display the trajectories at four different  relative strengths of incoming flow, $\tilde u_{in}=0, 1/4, 1/2, 1$, corresponding to incoming velocities $u_{in}=0, 5, 10, 20\,\mu m s^{-1}$, respectively. In (a) and (d), particles' trajectories distribute uniformly in the right and left sides of the channel, while for intermediate $Pe_f$, subplots (b),(c), the trajectories are more concentrated on the right side of the constriction, resulting in  asymmetric density profiles.}
		\label{f:traj1}
	\end{figure}
	In order to analyze the qualitative behavior of the system, we first consider SPPs trajectories in the channel as a function of the incoming fluid flow. Fig.~\ref{f:traj1} displays the results of 50  independent (non-interacting) particles for different values of relative incoming flow strengths, $\tilde u_{in}=0,1/4,1/2,1$, corresponding to incoming fluid velocities $u_{in}=0, 5,10, 20\, \mu m s^{-1}$. 
	
	For a  static fluid, Fig.~\ref{f:traj1}.a,  SPPs  are uniformly distributed throughout the channel, while for nonzero incoming fluid velocities, Fig.~\ref{f:traj1}.b-c, the right side of the constriction is more crowded. By increasing $\tilde u_{in}$, the asymmetry in the particle population  diminishes and finally disappears, as it can be seen in  subplot Fig.~\ref{f:traj1}.d. This non-monotonic behavior, which was first observed and reported  for dispersion of {\sl Ecoli} bacteria through a funnel~\cite{clement2013}, is a consequence of the interplay between  bacteria self-propulsion,  fluid flow and channel confinement. In the following sections we study the time evolution of SPPs' probability distribution function, their streamlines, and mean square displacement to reveal the underlying dynamics of this phenomenon. We will also discuss the controlling parameters which strengthen/weaken the observed symmetry breaking in population density before and after the constriction.  
	
	\subsection{Symmetry breaking in probability distribution}\label{s:SB}
	
\begin{figure}
		\centering
		\includegraphics[width=\linewidth]{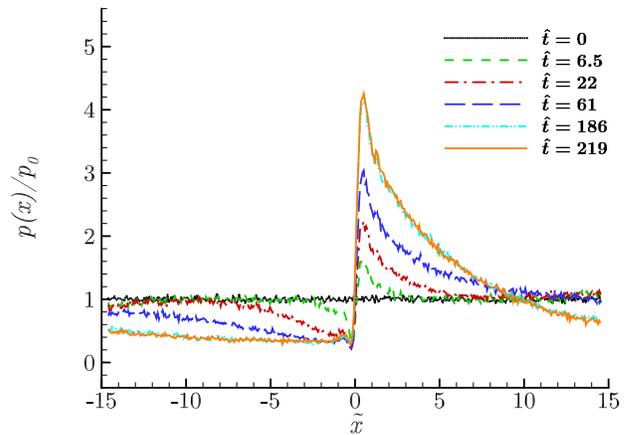}
		\caption{Time evolution of $p(x)$ for $N=4\times10^4$ ellipsoidal particles of aspect ratio $\lambda=6$ in the same channel as described in Fig~\ref{f:schematic}, averaged over 10 realizations. Starting from a uniformly distributed state (gray dotted line), the PDF evolves to an asymmetric distribution, for $t>200\tau_p$, with a sharp long-ranged peak just on the right side of the constriction (orange solid line).}
		\label{f:pdf-evo}
	\end{figure}

	In this section we study the time evolution of the probability distribution function (PDF), $p(x)$, starting from a uniform distribution with mean value $p_0=1/L_x$ in the  channel depicted in Fig.\ref{f:schematic}. The results for a system with $N=4\times10^4$ non-interacting ellipsoidal particles of aspect ratio, $\lambda=6$, $Pe_s=45.7$ and $\tilde u_{in}=1/4$ are shown in Fig.~\ref{f:pdf-evo}, where the dimensionless PDF, i.e.  $p(x)/p_0$, is averaged over 10 realizations.  The initial, uniform configuration (gray dotted line), evolves to a stationary asymmetric distribution for $t\geq200\tau_p$ (orange solid line). In this stationary state, an abrupt decrease in density  before the constriction, is followed by a sharp rise just after it, showing that particles are accumulated in the right side of the channel. This is in agreement with the dense trajectory lines  observed in Fig.~\ref{f:traj1}.b.
	
	Fig.~\ref{f:pdf-evo}  also shows a long range asymmetric decay of the PDF,  which extends  through the channel far beyond the constriction, consistent with experimental observations~\cite{clement2013}. It is also worth pointing out that $t_s\simeq200\tau_p$ is a safe enough choice to  ensure that the stationary state has been reached and time averaged quantities are well defined as is observed in other simulations with different parameters (data not shown here). 
	
	\begin{figure}
		\centering
		\includegraphics[width=\linewidth]{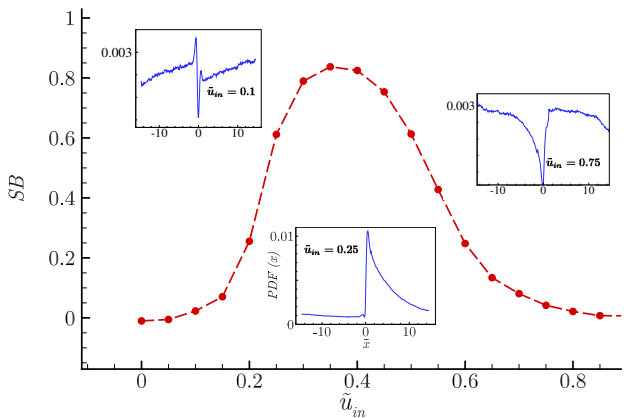}
		\caption{$SB$ as a function of relative incoming flow strength, $\tilde u_{in}$, for SPPs of aspect ratio $\lambda=6$. Symbols show simulated results, while the line is  a guide to the eye. Insets show PDF ($p(x)$) for three special cases, corresponding to  $\tilde u_{in}=1/10, 1/4, 3/4$.}
		\label{f:SB}
	\end{figure}
	
	We may now define a symmetry breaking parameter, $$SB=\frac{N_R-N_L}{N_R+N_L},$$ to quantify the observed unbalance between the population of particles in the left ($N_L=N\int_{-L_x/2}^{0} p(x) dx$) and right ($N_R=N\int_{0}^{+L_x/2} p(x) dx$) sides of the channel ~\cite{clement2013}.
	Fig.~\ref{f:SB} shows  $SB$  as a function of the relative incoming flow strength. Each  point is obtained by averaging over 300 snapshots. The previous expression assumes that the system reaches its stationary state, at $t=200\tau_p$. The time interval between two successive snapshots is $\sim0.07\tau_p$. In agreement with particles' trajectories,  $SB$ displays  a non-monotonic behavior in response to the increase of $\tilde u_{in}$.  We can distinguish three regimes: (i) zero symmetry breaking or homogeneous distribution for $\tilde u_{in}\simeq0$, (ii) nonzero finite symmetry breaking for middle relative flow strength  ($0.2\precsim \tilde u_{in}\precsim0.6$) and (iii) reduction to zero symmetry breaking (or uniform distribution) for large relative flow strengths $\tilde u_{in}\succsim0.75$. Therefore there exists a critical relative incoming flow strength, $\tilde u_{in}^*$, for which the system experiences the maximum symmetry breaking, $SB^*$.  
	
\begin{figure}
		\centering
		\includegraphics[width=\linewidth]{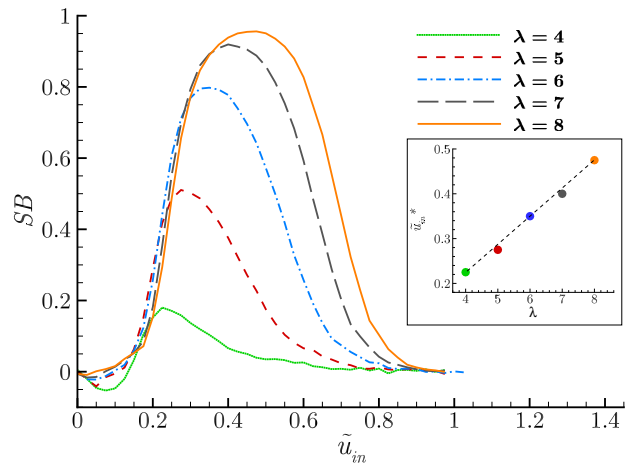} \\
		\caption{ $SB$, as a function of $\tilde u_{in}$ for ellipsoidal particles of various aspect ratios and fixed semi-minor axis $b=0.5\,\mu m$. In the insets, critical relative incoming flow strength, $\tilde u_{in}^*$ is plotted as a function of $\lambda$, indicating that the peak position increases linearly with particle asymmetry.}
		\label{f:SB-size}
	\end{figure}

\begin{figure}
		\centering
		\includegraphics[width=0.47\linewidth]{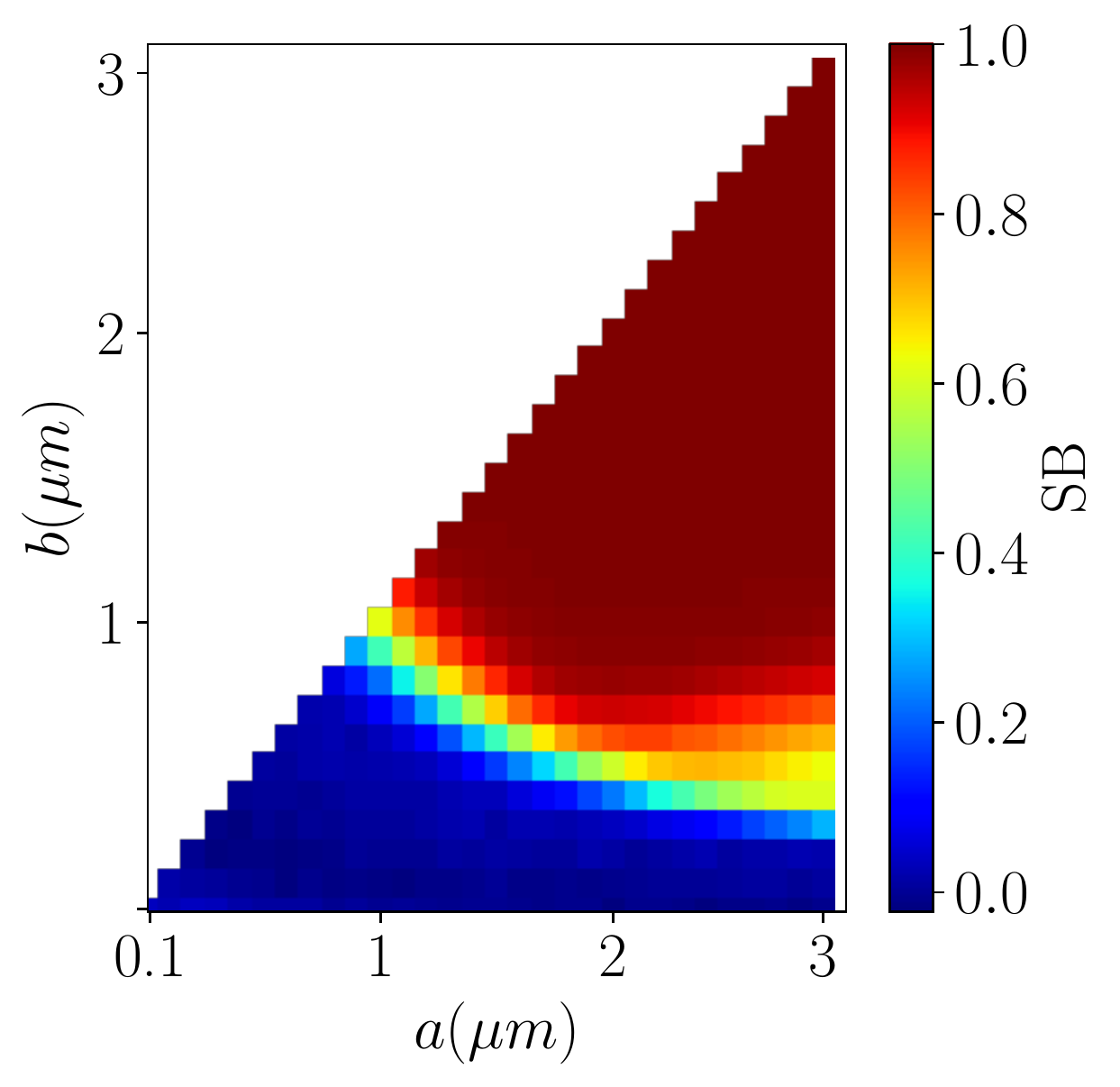} 
		\includegraphics[width=0.50\linewidth]{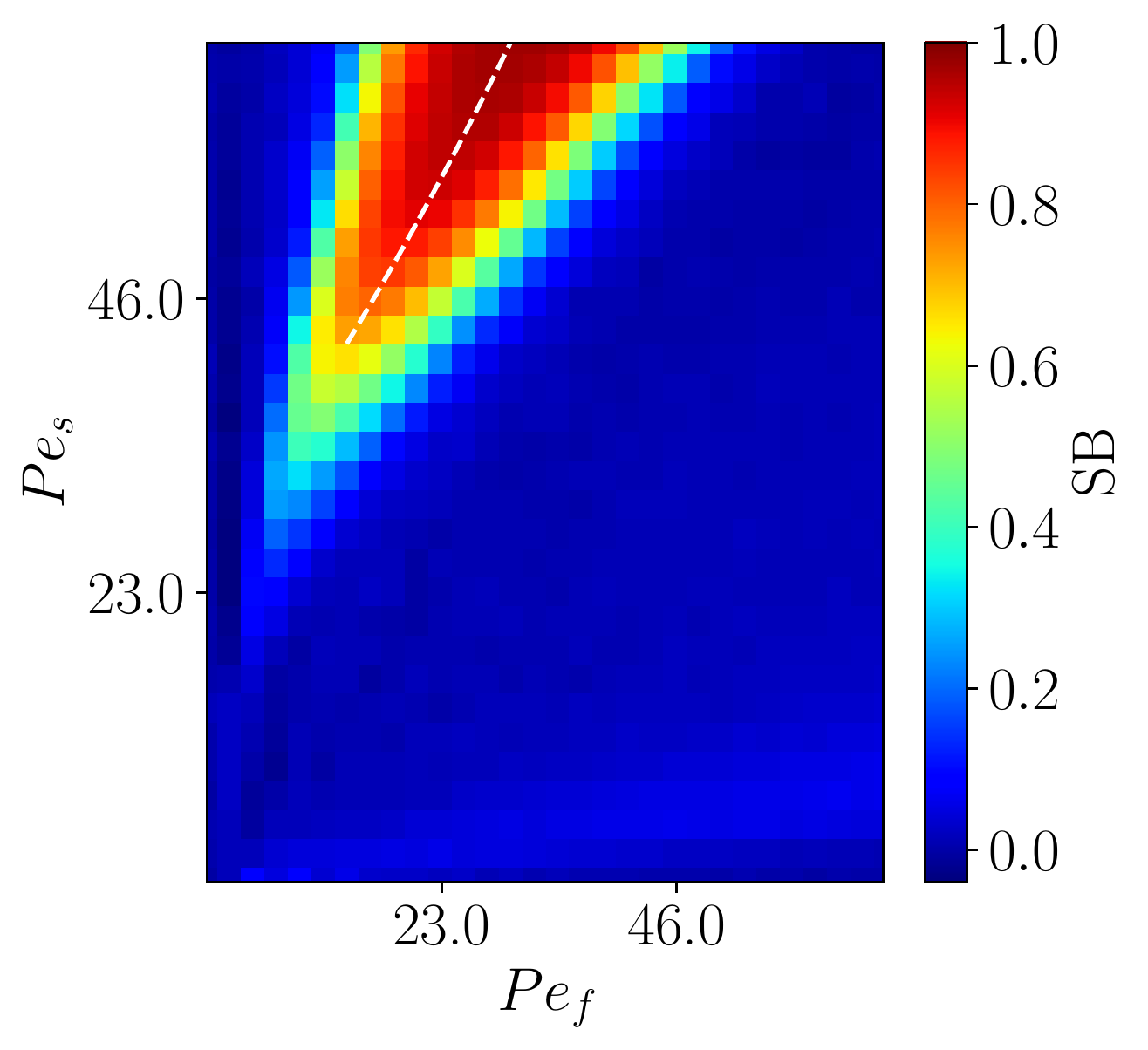}
		\caption{$SB$ a) as a function of SPP size and asymmetry for  $\tilde u_{in}=1/4$,  $Pe_s=45.7$, and b) as a function of the self-propelling and incoming velocities for a fixed particle shape, $\lambda=6$. The rest of the  parameters are the same as in Fig.~\ref{f:schematic}.}
		\label{f:SB-phase}
	\end{figure}

	The $SB$ profile as well as the critical values ($\tilde u_{in}^*$ and $SB^*$) are influenced by various factors including particle asymmetry, relative strength of the incoming flow with respect to self-propulsion speed, and channel confinement. Fig.~\ref{f:SB-size},  indicates that both $SB$ and $\tilde u_{in}^*$ (inset) generically grow with particle elongation, $\lambda$. Fig.~\ref{f:SB-phase} provides a more comprehensive view of the impact that SPP geometry and incoming flow has on the asymmetric organization of SPPs around the channel constriction. Such a global view is insightful in view of the  large disparity of size and shape of artificial and natural SPPs, which range from sphere to needle-like, as well as the different velocities at which they self-propel. Specifically, Fig.~\ref{f:SB-phase}.a shows that not  only the aspect ratio, but also the particle geometry (combination of radii) could influence the value of $SB$. In the right lower corner of this figure, for instance, we find needle-like particles with large aspect ratio and small $SB$ (almost zero), in contrast to the general trend.
	
\begin{figure}
		\includegraphics[width=\linewidth]{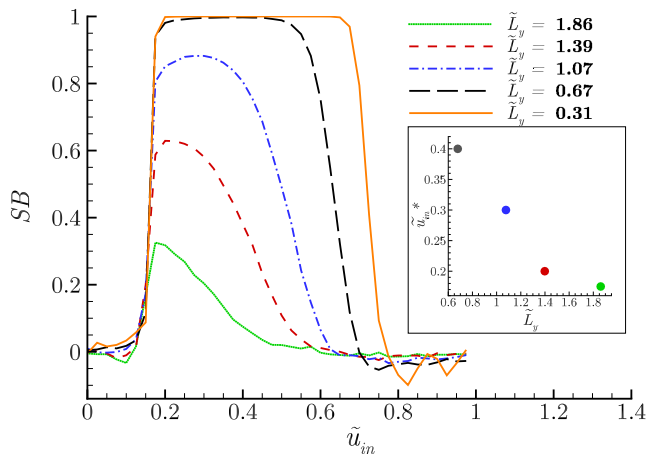}\\
		\caption{Symmetry breaking versus relative incoming flow strength for various confinement parameters ($lp/L_y=\tilde L_y^{-1}$). Inset shows how the maximum value of SB profile decrease by decreasing confinement parameter (or increasing $\tilde L_y$).}
		\label{f:SB-conf}	
	\end{figure}

	Fig.~\ref{f:SB-phase}.b  shows the relevant role played by the relative magnitudes of the self propelling and incoming fluid velocities. For a fixed $Pe_s$, $SB$ increases  from zero to its maximum value and back to zero by increasing $Pe_f$, in agreement with Fig.~\ref{f:SB}. For larger propulsion speeds (e.g. $Pe_s\simeq70$), symmetry breaking occurs for a wider range of flow velocities, such that the system goes sharply from zero to its maximum value and form a plateau rather than a single maximum point (This is similar to the situation observed for high persistent length (or low $\tilde L_y$) in Fig. \ref{f:SB-conf}). The other way around, for a fixed value of $Pe_f$, $SB$ increases to its maximum value by increasing $Pe_s$, and returns back to symmetric state when $Pe_s$ becomes large enough. 
	
	It is worth noting that the states with equal SB lie on two crossing lines in velocity phase space, which first confirms the existence of a maximum point in SB profile either by changing the fluid or propulsion velocities, and second indicates that there is a unique linear relation between the two velocities for all peak points of SB profile; i.e. all the peak points for various propulsion speeds lie on the same line in velocity phase space; here indicated by white dashed line: $Pe_s^*=1.5 Pe_f^*+9.7/(a D_R)$.
	
	The impact of confinement, on the other hand, is characterized by confinement parameter, $l_p/L_y=\tilde L_y^{-1}$. The distribution of particles in the channel becomes more asymmetric by decreasing $\tilde L_y$, as is indicated in Fig.~\ref{f:SB-conf}. 
	Note that single maximum point in low confinement regime ($\tilde L_y>1$) changes to a plateau in highly confined ($\tilde L_y<1$) cases. 
	
	The negative $SB$ region, observed in Figs.\ref{f:SB-size}, \ref{f:SB-conf} for  weak relative fluid strength ($\tilde u_{in}\precsim 1/10$), is due to the fact that particles are blocked behind the constriction, then begin to pass it and aggregate on the right side by increasing the flow velocity. 
	
	\subsection{Particles' streamlines  and upstream flow} 
	
	\begin{figure}
		\centering 
		\includegraphics[width=1.0\linewidth]{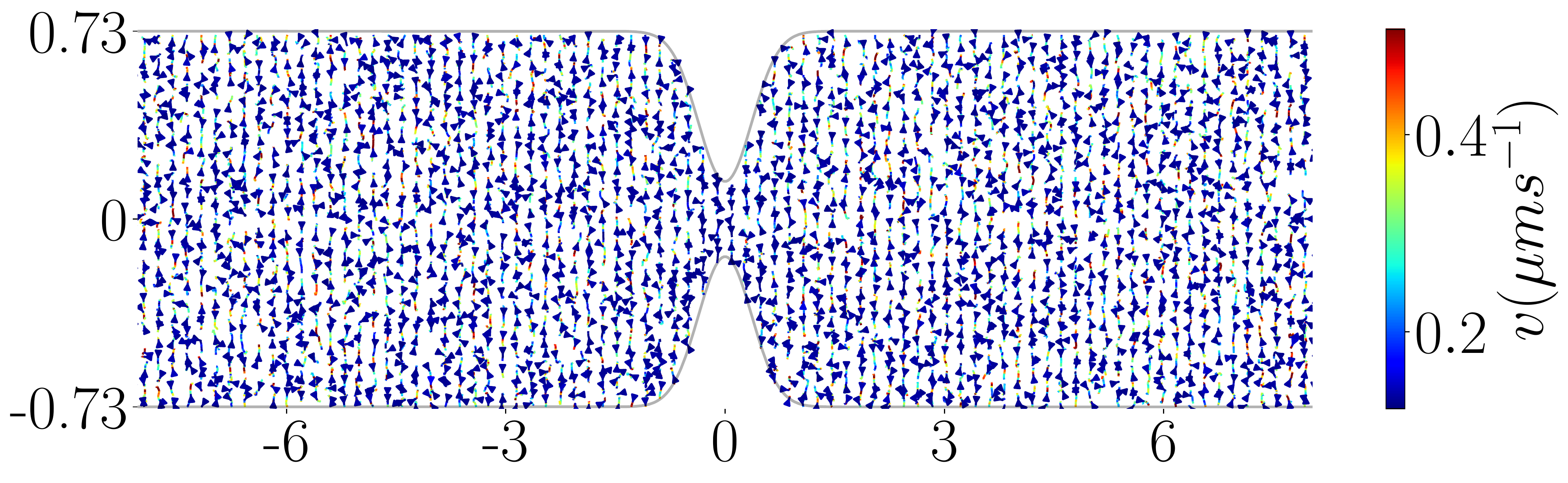} (a) $\tilde u_{in}=0$\\
		\includegraphics[width=1.0\linewidth]{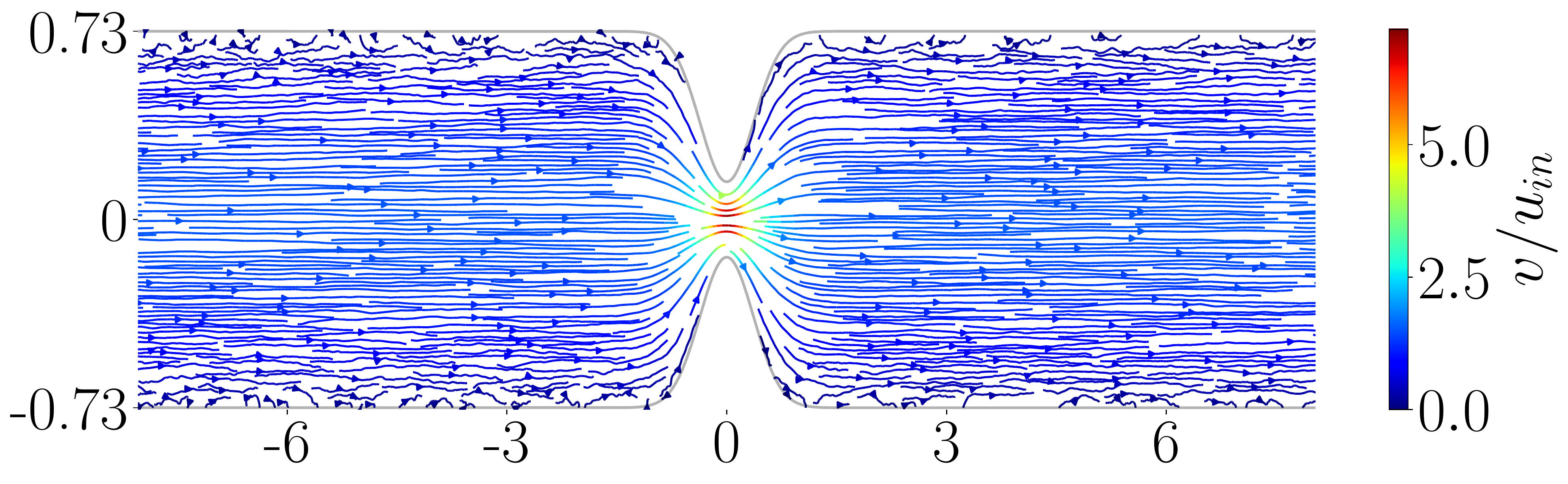} (b) $\tilde u_{in}=1/4$\\
		\caption{Streamlines of passive particles ($v_s=0$) moving in the channel of width $\tilde L_y=1.46, \tilde L_c=0.29$, for relative incoming flow strengths $\tilde u_{in}=0, 1/4$. The velocity vector field, $\mathbf v(x,y)$, is the local average velocity of particles found in surface element $\delta x \delta y$ around each point, $(x,y)$, in the channel. The color of streamlines changes by the absolute value of average velocity from zero (blue) to its maximum value (red) as is shown in the color bar. The triangles on the streamlines show the direction of average velocity field.}	
		\label{f:ps-strplot}
	\end{figure}
	
	\begin{figure}
		\centering 
		\includegraphics[width=1.0\linewidth]{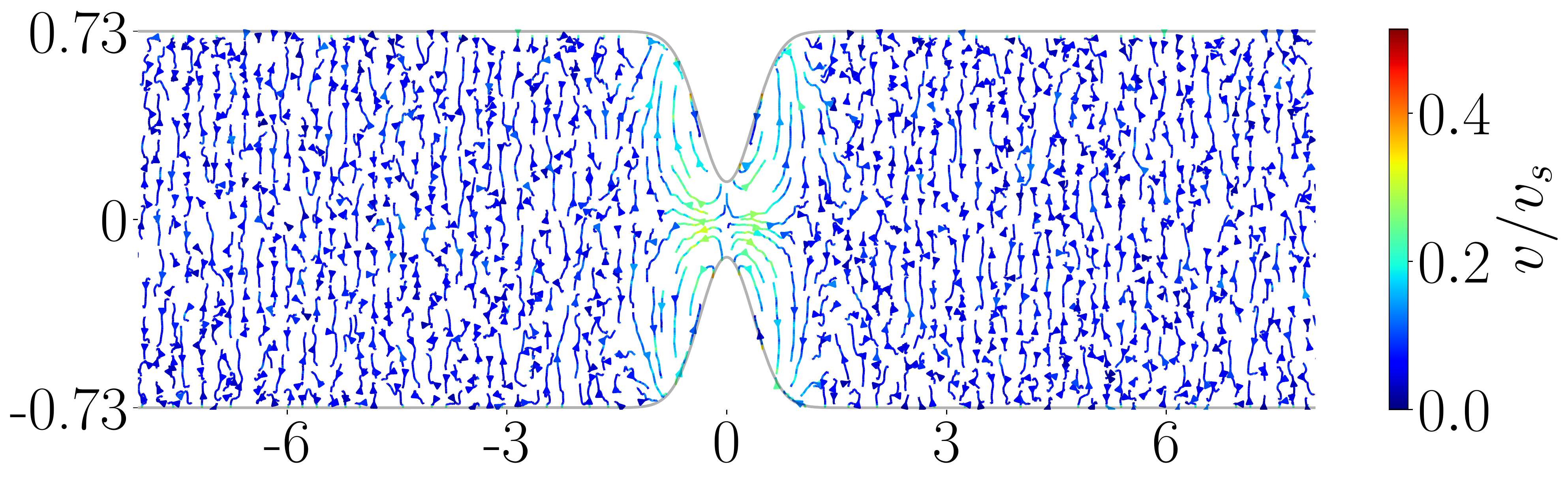} (a) $\tilde u_{in} = 0$\\
		\includegraphics[width=1.0\linewidth]{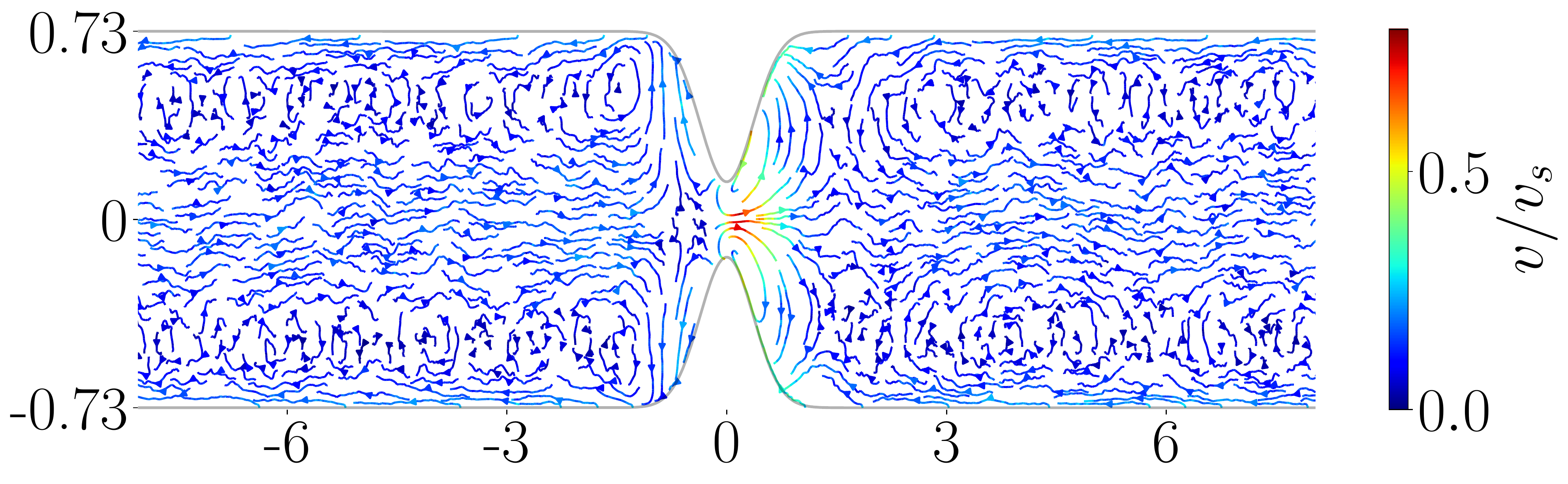} (b) $\tilde u_{in}=1/10$\\
		\includegraphics[width=1.0\linewidth]{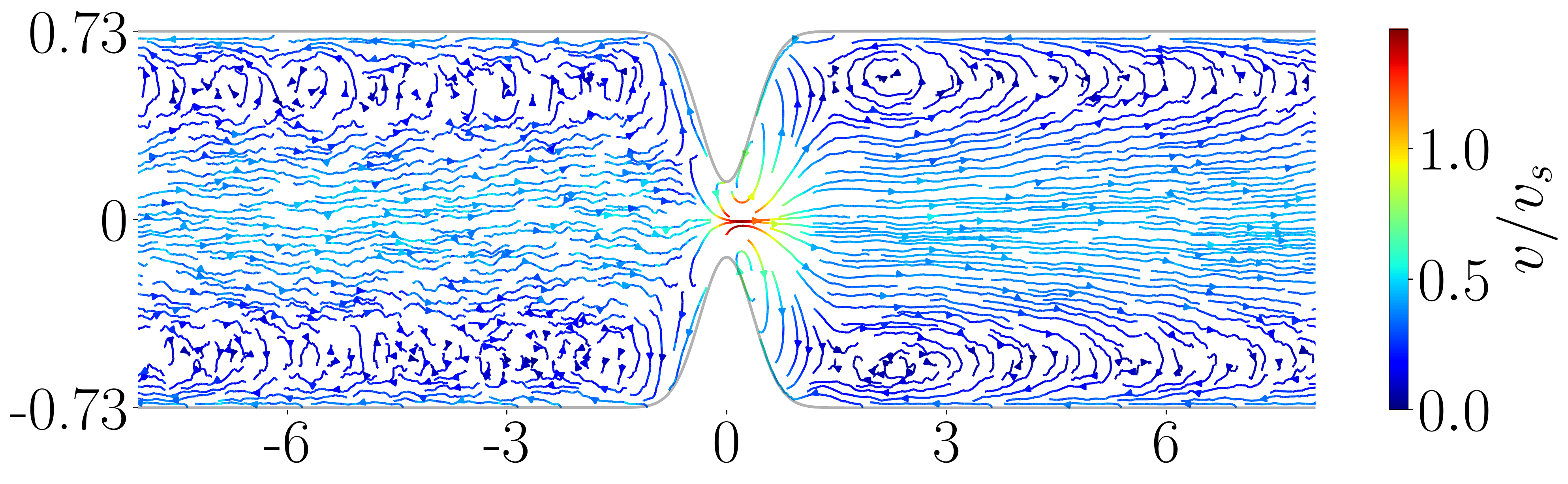} (c) $\tilde u_{in}=1/4$\\
		\includegraphics[width=1.0\linewidth]{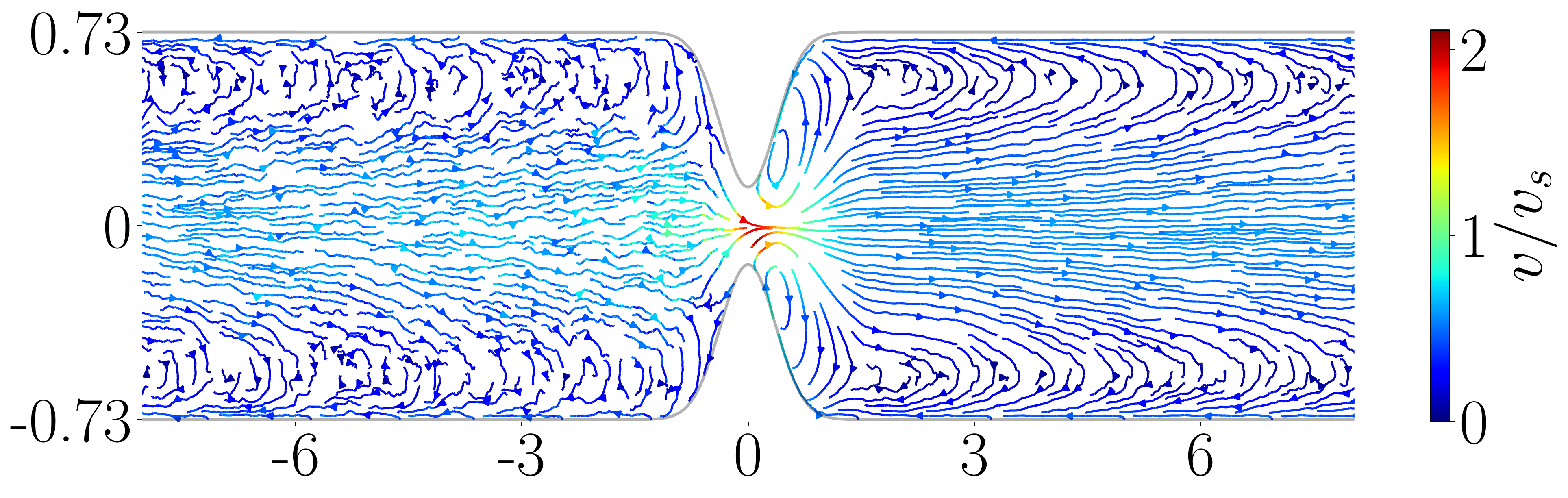} (d) $\tilde u_{in}=3/10$\\
		\includegraphics[width=1.0\linewidth]{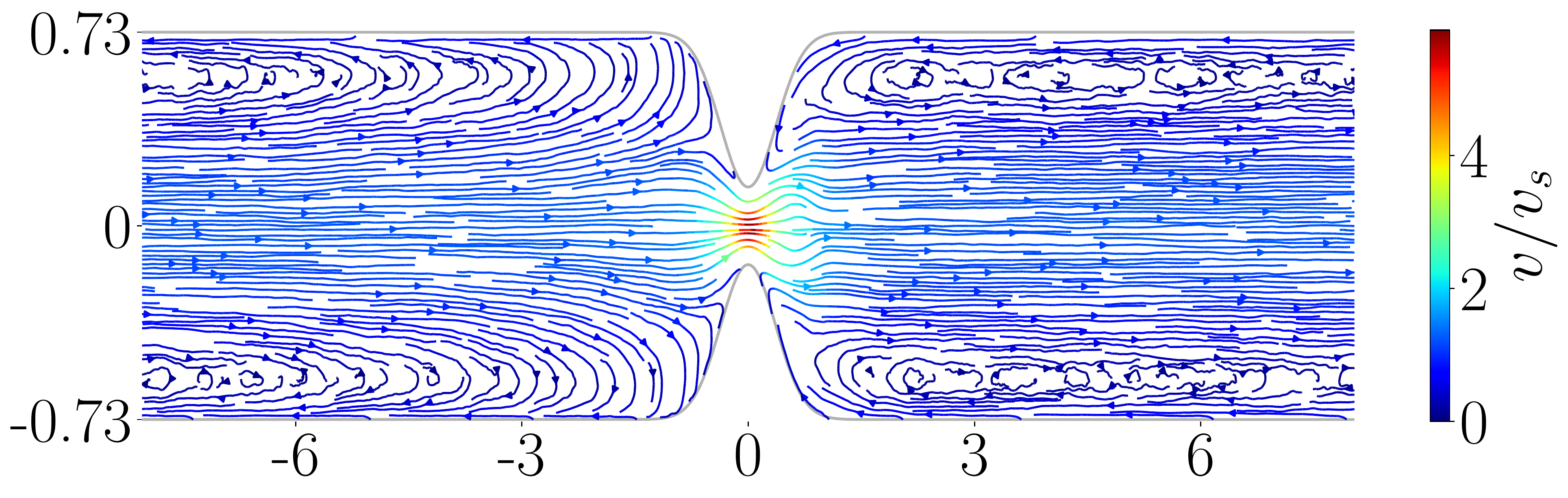} (e) $\tilde u_{in}=3/4$\\
		\caption{Streamlines of SPPs of aspect ratio $\lambda=6$ moving in the same channel as described in Fig.~\ref{f:ps-strplot} for $\tilde u_{in}=0, 1/10, 1/4, 3/10,3/4$. Vertical/horizontal axes are rescaled by $\l_p$.}
		\label{f:strplot}
	\end{figure}
	
	\begin{figure}
		\centering 
		\includegraphics[width=\linewidth]{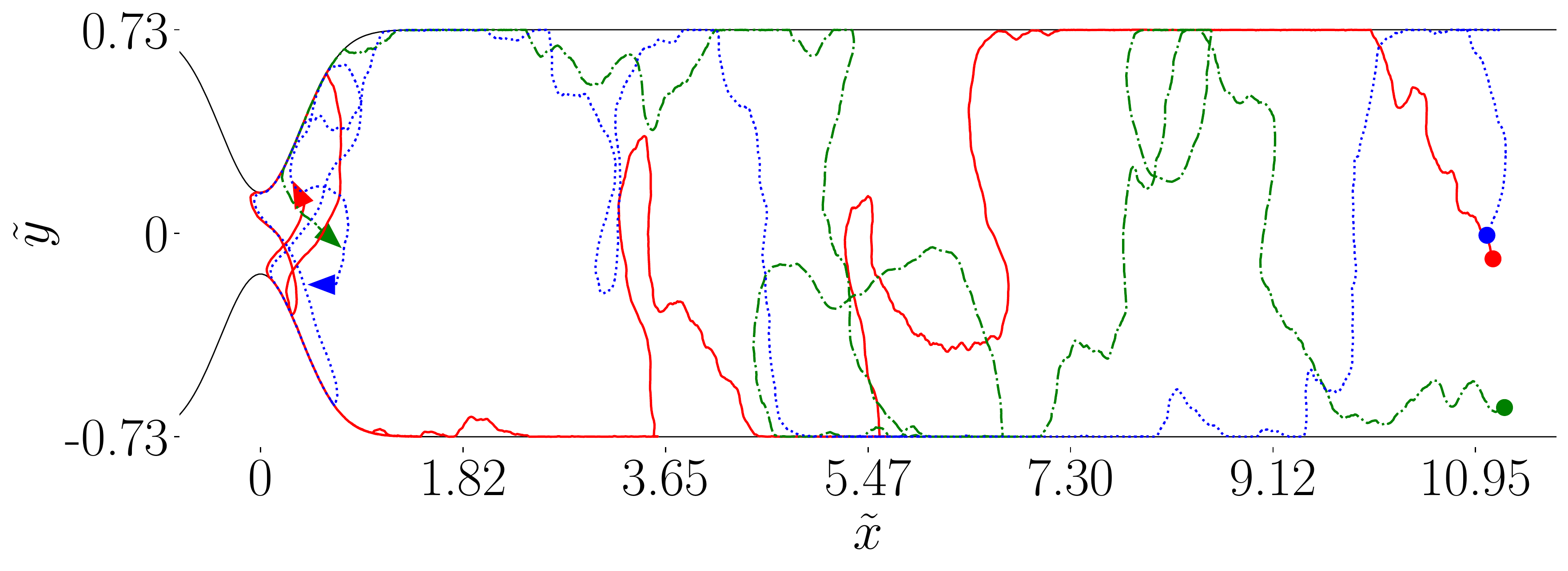} \\
		\caption{Typical trajectories of SPPs ($\lambda=6$) moving upstream in the same channel as described in Fig.~\ref{f:ps-strplot}., for$Pe_f=0.3$, corresponding to maximum symmetry breaking. Starting from the right side of the channel (filled circles), particles experience an up-and-down motion between the walls and finally are trapped in a butterfly-like trajectory in front of the constriction. Arrows show the direction of particles' velocity at the ending point of their trajectories.}
		\label{f:traj3}
	\end{figure}
	
	In order to understand the  types of  motion that lead to symmetry breaking, we   analyze the SPPs streamlines (i.e. local direction of their velocity field). The velocity field, $\mathbf v(x,y)$, is obtained by measuring the local average velocity of particles found in surface element $\delta x \delta y$ around the point $(x,y)$, in the channel. In the stream plots, the value and direction of average velocity are shown by colored vectors which, according to the color code given in the side bars, changes from zero velocity (blue) to their maximum magnitude (red). We first analyze the special case of passive particles and then discuss  activity-induced effects. 
	
	\paragraph{Passive particles--} Fig.~\ref{f:ps-strplot} displays passive particles ($v_s=0$) moving in the  channel of Fig~\ref{f:schematic} with $\tilde u_{in}=0, 1/4$.  As  expected, for a fluid at rest, Fig.~\ref{f:ps-strplot}.a,  particles  move uniformly in all directions and do not show any preferred direction in their motion. For a net incoming fluid flow, Fig.~\ref{f:ps-strplot}.b, particles are advected along the flow direction throughout the channel (downstream flow). The local velocity equals the incoming velocity on the left/right walls and is zero on the upper and lower walls as is set by the boundary conditions. Comparing to Fig.~\ref{f:schematic}.b,  the local average value of particles' velocity is the same as that of the fluid, indicating that, in agreement with analytical results, the Brownian term in the equations of motion of passive particles is averaged out and the leading term is the fluid velocity. Therefore,  passive particles either display  Brownian motion, at $u_{in}=0$,  or statistically follow the fluid flow.
	
	\paragraph{Active particles--} Fig.~\ref{f:strplot} displays the streamlines for SPPs of constant P\'{e}clet, $Pe_s=45.7$, for a fixed channel geometry and increasing incoming fluid flows, $\tilde u_{in}=0, 1/10, 1/4, 3/10, 3/4$. For a quiescent liquid, Fig.~\ref{f:strplot}.a,  the streamlines do not show a preferred direction and the average velocity  essentially  vanishes, except in vicinity of the constriction where the velocity is above average and particle self propulsion form a symmetric butterfly pattern with its wings extended around the constriction. By comparing with the streamlines of passive particles, Fig.\ref{f:ps-strplot}(a), it is apparent that the butterfly pattern is a consequence of the particles' self propulsion in the presence of a constriction.
	
	For nonzero incoming fluid velocities, Fig.~\ref{f:strplot}.b-e, we  observe aligned streamlines in the center of the channel as  expected.  For intermediate incoming  fluid velocities, Fig.~\ref{f:strplot}.b-d, the  butterfly wings observed for a quiescent liquid become progressively more asymmetric,  such that for the critical flow velocity where SB is maximized, Fig.~\ref{f:strplot}.d, the butterfly left wings completely disappear. By further increasing the fluid velocity, Fig.~\ref{f:strplot}.e, the butterfly pattern is replaced  by aligned streamlines indicating that advective motion is dominant..
	
	Comparing the streamline patterns in Fig.~\ref{f:strplot} with the trajectories in Fig.~\ref{f:traj1}, it can be concluded that the right wings of the butterfly pattern are  actually the region where particles are trapped in front of the constriction. This type of attractor is also reported in the motion of sperm cells passing a funnel~\cite{sperm}. The extra time spent in the right side of the channel, for net incoming flow, is actually the origin of the observed asymmetric distribution of particles ~\cite{clement2013}. For low fluid P\'{e}clet numbers, $\tilde u_{in}\precsim 1/5$,  asymmetric butterfly wings do not develop, while for high P\'{e}clet numbers, $\tilde u_{in}\succsim 3/4$, the downstream convective motion is dominant; hence  trapping is not possible  in these two limiting regimes. The maximum symmetry breaking is therefore observed for intermediate incoming fluid velocities ($1/5\precsim \tilde u_{in} \precsim 3/5$) where the strength of these opposing effects are balanced.
	
	Upstream flow, {\sl i.e.} moving in the opposite direction of the flow, is another characteristic feature of SPPs that can reinforce/support  symmetry breaking. In contrast to downstream flow, which is the expected motion in the direction of the flow, upstream flow provides the chance for particles that have passed the constriction, to move back towards the center of the channel and be trapped in the  wings of the butterfly pattern. Fig.~\ref{f:traj3} displays typical particle trajectories  starting  on the right side of the channel, moving upstream,  finally  trapped in front of the constriction. Therefore, the trapping probability and thus the value of $SB$  as well as the value of critical velocity, are correlated with the ability of particles to perform  upstream flow, which itself is  affected by the ratio of self propulsion speed to fluid velocity,
	the size/shape of particles or geometry of boundaries. 
	
	Particles' streamlines also develop vortices near the upper and lower walls, in which one can easily distinguish the upstream flow in the very first layers close to the walls. The emergence of vortices is due to the interplay of self propulsion and fluid flow; they disappear both for passive particles moving in a nonzero fluid flow, Fig.~\ref{f:ps-strplot}, and active particles in stationary flow, Fig.~\ref{f:strplot}(a).
	
	\subsection{Mean  Square Displacement}
	\begin{figure} 
		\centering
		\includegraphics[width=\linewidth]{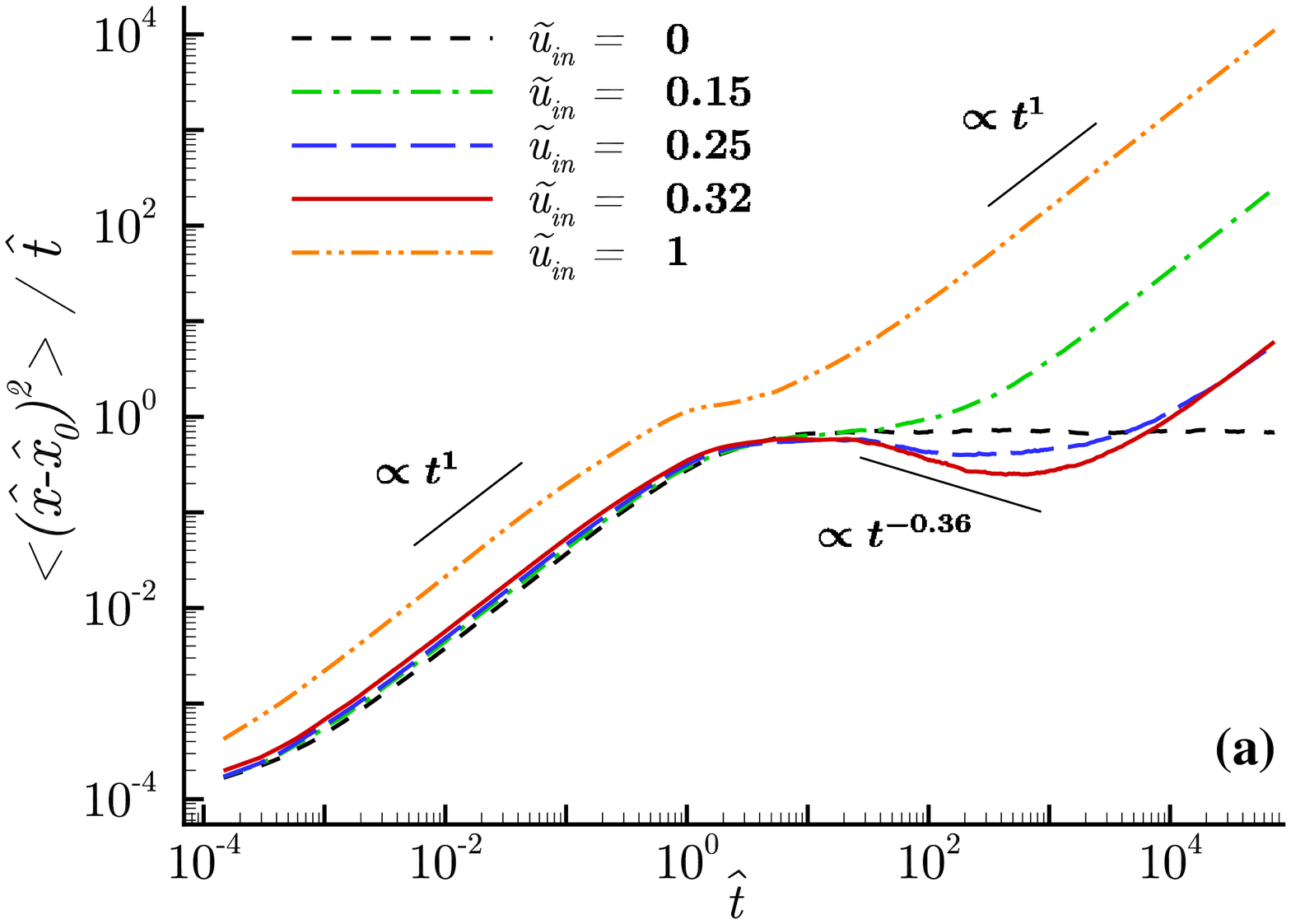} 
		\includegraphics[width=\linewidth]{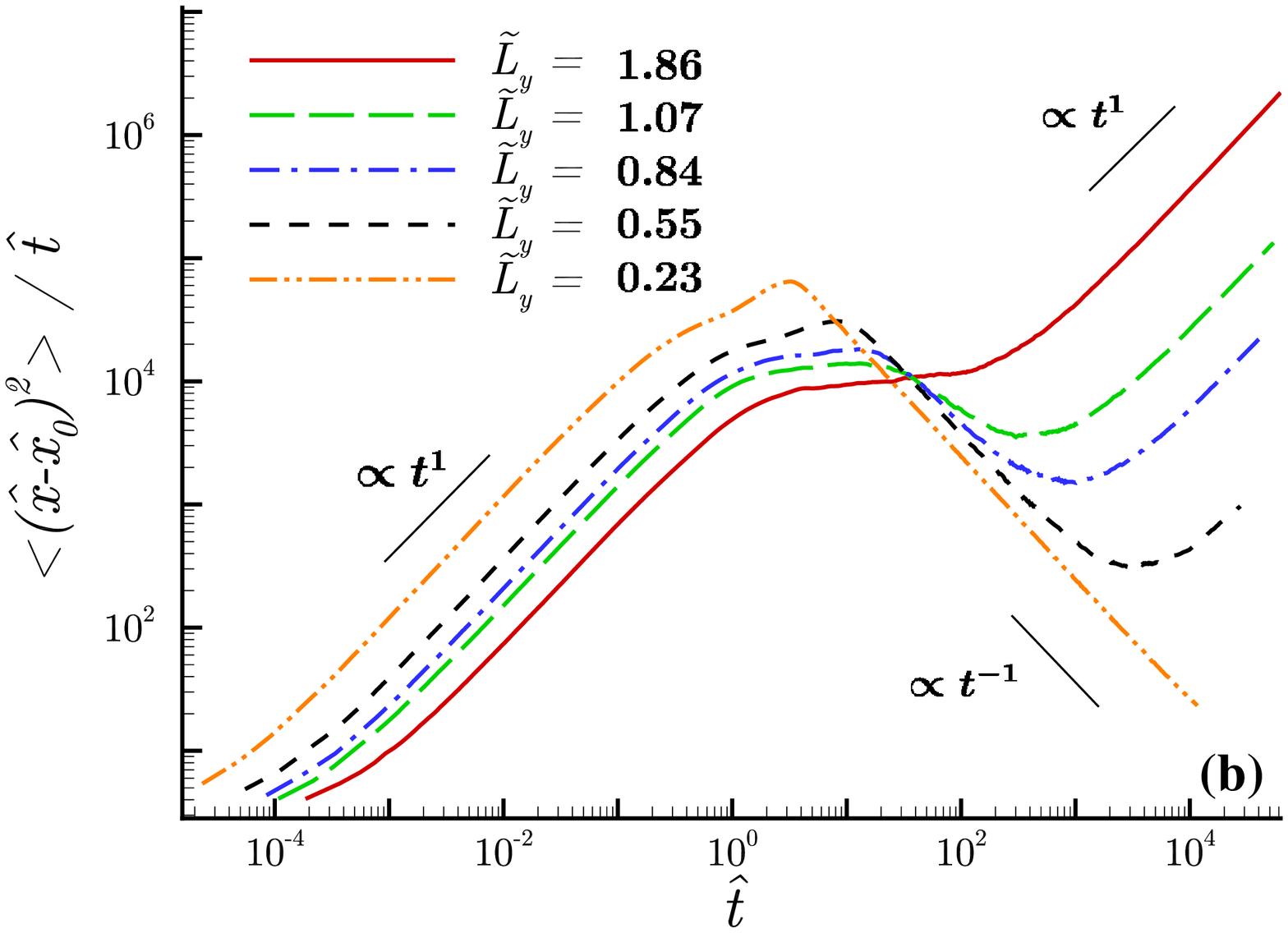} 
		\caption{Mean square horizontal displacement (MSD) divided by time for (a) ellipsoidal particles of aspect ratio $\lambda=6$ for different incoming flow $\tilde u_{in}$, and (b) spherical particles of different $\tilde L_y$ (corresponding to radii $a=1 ,1.2, 1.3, 1.5, 2.0 \mu m$, respectively) and $\tilde u_{in}=1/4$.}
		\label{f:msd}	
	\end{figure}
	
	The mean square displacement (MSD) of SPPs provides additional insight into their effective dynamics. The MSD  is affected by  background fluid flows and/or complexities of the environment, such that multiple transitions from ballistic to diffusive, and sub(or super)-diffusive  regimes might be observed. 
	
	In the absence of  an incoming flow, the MSD of  SPPs is characterized by a ballistic and an asymptotic diffusive regimes  with a crossover time scale  determined by  rotational diffusion. Typically, $\tau_D=1/D_R\simeq  1-10 s$ for microorganisms and self propelling colloids, is much larger than the time scales in which passive colloidal particles of mass $m$ and friction coefficient, $\gamma$, enter the diffusive regime, $t_D=m/\gamma\simeq 10^{-9}$ s.
	
	Fig.~\ref{f:msd} displays the MSD divided by time in dimensionless units for various  (a)  $\tilde u_{in}$, and (b)  $\tilde L_y$. 
	
	In the absence of an incoming flow, Fig.~\ref{f:msd}(a), the MSD crosses over from ballistic to normal diffusive regime at $\tau_D\simeq10 s$ (black dashed line). For nonzero incoming fluid velocities, the diffusive regime lasts for a finite time and, asymptotically, the  MSD shows another crossing to an asymptotic ballistic regime; which is now a consequence of particle advection by the fluid flow. These three regimes are  clearly distinguished  for low ($\tilde u_{in}=3/20$) and high ($\tilde u_{in}=1$) incoming fluid strengths; while for  intermediate values ($\tilde u_{in}= 1/4, 6.5/20$), we observe an extra  sub-diffusive regime from $\hat t\simeq 40$ to $4000$, followed by a ballistic regime, which in this case is a very weak advection even compared to the lowest velocity (i.e. $\tilde u_{in}= 3/20$). This characteristic feature of intermediate velocities confirms the trapping of particles in the channel (i.e. butterfly trajectories), which is the origin of the observed symmetry breaking. 
	
	By increasing the persistent length (or decreasing $\tilde L_y$), as displayed in Fig.~\ref{f:msd}.b, the duration of the sub-diffusive regime increases and the final advective motion becomes weaker (note the region with negative slope) in agreement with the increase of symmetry breaking indicated in Fig.~\ref{f:SB-conf}. The MSD for large persistent lengths (e.g. $\tilde L_y=0.23$) goes asymptotically to  a constant value. It eventually,  becomes ballistic but at much longer times (not shown here); meaning that a high percentage of particles are almost motionless, leading to high symmetry breaking ($SB\simeq1$) for wide range of velocities as is also seen in Fig.~\ref{f:SB-conf}.
	
	\section{Conclusions}
	\label{sec:conclusions}
	We have performed Brownian dynamics simulations to investigate the underlying dynamics that lead to the  observed densification of self-propelled particles in confined micro-channels with a constriction~\cite{clement2013}. The model system considers the general case of self-propelling particles of arbitrary shape/size whose motion are simulated as active Brownian particles.  The propelling objects in this model perform an overdamped motion under the action of an internal driving force directed along its orientation, which itself undergoes Brownian fluctuations. In addition, both the translational and rotational degrees of freedom are affected by the nonlinear fluid velocity field in the channel, which is obtained by solving the Stokes equation, and assumed to be independent of the propelling objects.  
	
	The simulation results indicate that in  the stationary state, the particle  probability distribution in the channel can be either  symmetric or asymmetric  with respect to the constriction depending on the value of fluid P\'{e}clet number (or the ratio of fluid velocity to self-propulsion speed). The symmetry breaking parameter ($SB$), which quantifies the degree of asymmetry, is maximized for intermediate incoming flows, $Pe_f=3/10$, while it goes to zero asymptotically both for low and high fluid P\'{e}clet numbers, in agreement with  experimental evidence~\cite{clement2013}. 
	
	The critical average fluid velocity (with maximum $SB$) is linearly proportional to the propelling  speed of active particles; such that all peak points of SB profile lie on the same line in velocity phase space as shown in Fig.~\ref{f:SB-phase}. For constant propulsion and fluid velocities, the peak position ($Pe_f^*$) scales with particles' size, and its height (peak value) increases by increasing the size of particles and/or confinement.
	
	A closer analysis of the particles' streamlines has revealed that for intermediate fluid velocities, $1/5<Pe_f<3/5$, particles are trapped in butterfly-like trajectories at the constriction. This pattern traps the self-propelling particles preferentially when they meet the constriction through upstreaming leading to the observed symmetry breaking. The butterfly pattern develops due to the competition between the incoming flow and particle self-propulsion. The attractor, the butterfly pattern, is sustained for intermediate strengths of the incoming flow and depends on, among other parameters, the ratio of propulsion to flow speed and persistent length.
	The trace of the  attractors in MSD plots, appears as sub-diffusive regimes, whose duration increases for larger particles, in agreement with the larger symmetry breaking observed for larger particles.
	
	The model we have introduced is versatile, and can be applied to more complex boundary conditions, different channel geometries and/or other models of propulsion such as run-and-tumble motion. Building on the reported analogous dynamical behavior of ABP and run and tumble~\cite{EPL_CatesTaileur}, the good agreement with experimental results for E.Coli suggest that the ABP model for an ellipsoidal particle is able to capture  the essential features of the  response of E-Coli in constricted channels. The results reported show that symmetry breaking in confined channels with constriction is a generic property of self propelling particles, not restricted to special species, or sizes.
	
	\begin{acknowledgments}	
		This work is supported by Tarbiat Modares Univversity. I.P. acknowledges support from Ministerio de Ciencia, Innovacion y Universidades MCIU/AEI/FEDER for financial support under grant agreement PID2021-126570NB-100 AEI/FEDER-Eu, from Generalitat de Catalunya under Program Icrea Acad\`emia and project 2021SGR-673. 
	\end{acknowledgments}

	\bibliography{ref.bib}
\end{document}